\documentclass[traditabstract, longauth]{aa}

\usepackage{amsfonts}
\usepackage{amsmath}
\usepackage{amssymb}
\usepackage{pdflscape}
\usepackage{rotating}
\usepackage{natbib}
\usepackage{txfonts}
\usepackage{longtable}
\usepackage[small,bf]{caption}
\usepackage{subfigure}
\usepackage{multirow}
\usepackage{xspace}
\usepackage[pdftitle={G.Piano-Cyg-X3-V15}, colorlinks=true, citecolor=blue, linkcolor=blue, urlcolor=blue, pdfauthor={Giovanni Piano}]{hyperref}

\newcommand{\de}{\mathrm{d}}

\renewcommand{\rho}{\varrho}

\def \gray {$\gamma$-ray\xspace}
\def \grays {$\gamma$-rays\xspace}
\def \flx {photons $\mathrm{cm}^{-2}$ $\mathrm{s}^{-1}$\xspace}
\def \grid {AGILE-\textit{GRID}\xspace}

\def \lat {\textit{Fermi}-LAT\xspace}

\def \hp {1AGL J2021+3652\xspace}
\def \gcyg {1AGL J2022+4032\xspace}
\def \srcyg {1AGL J2032+4102\xspace}
\def \hpsr {PSR J2021+3651\xspace}
\def \gcygpsr {PSR J2021+4026\xspace}
\def \srcygpsr {PSR J2032+4127\xspace}

\begin{document}

\title{The AGILE monitoring of Cygnus X-3: Transient gamma-ray emission and spectral constraints}
\titlerunning{--}

\author{
G.~Piano\inst{1,2} \and
M.~Tavani\inst{1,2,3,8} \and
V.~Vittorini\inst{1} \and
A.~Trois\inst{9} \and
A.~Giuliani\inst{4} \and
A.~Bulgarelli\inst{5} \and
Y.~Evangelista\inst{1} \and
P.~Coppi\inst{15} \and
E.~Del Monte\inst{1} \and
S.~Sabatini\inst{1,2,8} \and
E.~Striani\inst{3,8} \and
I.~Donnarumma\inst{1} \and
D.~Hannikainen\inst{16,17} \and
K.~I.~I.~Koljonen\inst{16} \and
M.~McCollough\inst{18} \and
G.~Pooley\inst{19} \and
S.~Trushkin\inst{20} \and
R.~Zanin\inst{21} \and
G.~Barbiellini\inst{6} \and
M.~Cardillo\inst{1,3} \and
P.~W.~Cattaneo\inst{7} \and
A.~W.~Chen\inst{4} \and
S.~Colafrancesco\inst{12,13} \and
M.~Feroci\inst{1} \and
F.~Fuschino\inst{5} \and
M.~Giusti\inst{1,2} \and
F.~Longo\inst{6} \and
A.~Morselli\inst{8} \and
A.~Pellizzoni\inst{9} \and
C.~Pittori\inst{11,12} \and
G.~Pucella\inst{10} \and
M.~Rapisarda\inst{10} \and
A.~Rappoldi\inst{7} \and
P.~Soffitta\inst{1} \and
M.~Trifoglio\inst{5} \and
S.~Vercellone\inst{14} \and
F.~Verrecchia\inst{11,12}
\\}

\institute{INAF/IAPS, via del Fosso del Cavaliere 100, I-00133 Roma, Italy     
\and CIFS-Torino, viale Settimio Severo 3, I-10133 Torino, Italy            
\and Dipartimento di Fisica, Universit\`a di Roma ``Tor Vergata'', via della Ricerca Scientifica 1,I-00133 Roma, Italy      
\and INAF/IASF-Milano, via E. Bassini 15, I-20133 Milano, Italy       
\and INAF/IASF-Bologna, via Gobetti 101, 40129 Bologna, I-40129 Bologna, Italy     
\and Dipartimento di Fisica and INFN Trieste, via Valerio 2, I-34127 Trieste, Italy      
\and INFN-Pavia, via Bassi 6, I-27100 Pavia, Italy     
\and INFN-Roma ``Tor Vergata'', via della Ricerca Scientifica 1, I-00133 Roma, Italy    
\and INAF-Osservatorio Astronomico di Cagliari, localit\`a Poggio dei Pini, strada 54, I-09012 Capoterra, Italy  
\and ENEA Frascati, via E. Fermi 45, I-00044 Frascati (Roma), Italy    
\and ASI Science Data Center (ASDC), via G. Galilei, I-00044 Frascati (Roma), Italy  
\and INAF-OAR, I-00040, Via Frascati 33, Monte Porzio Catone, Italy  
\and University of the Witwatersrand, School of Physics, WITS 2050 Johannesburg (South Africa) 
\and INAF-IASF-Palermo, via U. La Malfa 15, I-90146 Palermo, Italy  
\and Department of Astronomy, Yale University, P. O. Box 208101, New Haven, CT 06520-8101, USA  
\and Aalto University Mets\"{a}hovi Radio Observatory, Mets\"{a}hovintie 114 FIN-02540 Kylm\"{a}l\"{a}, Finland  
\and Department of Physics and Space Sciences, Florida Institute of Technology, 150 W. University Blvd., Melbourne, FL 32901, USA 
\and Smithsonian Astrophysical Observatory, 60 Garden Street, Cambridge, Massachusetts 02138, USA  
\and Astrophysics Group, Cavendish Laboratory, 19 J. J. Thomson Avenue, Cambridge CB3 0HE, UK  
\and Special Astrophysical Observatory RAS, Karachaevo-Cherkassian Republic, Nizhnij Arkhyz 369169, Russia  
\and Departament d'Astronomia i Meteorologia, Institut de Ci\`encies del Cosmos, Facultat de F\'isica, 7a planta, Universitat de Barcelona
Mart\'i i Franqu\`es 1, 08028 Barcelona, Spain  
}

\abstract
{We present the \grid (Astro-rivelatore Gamma a Immagini LEggero -- \textit{Gamma-Ray Imaging Detector}) monitoring of \mbox{Cygnus X-3}, during the period between November 2007 and July 2009. We report here the whole \grid monitoring of \mbox{Cygnus X-3} in the AGILE ``pointing'' mode data-taking, to confirm that the \gray activity coincides with the same repetitive pattern of multiwavelength emission and analyze in depth the overall \gray spectrum by assuming both leptonic and hadronic scenarios. Seven intense \gray events were detected in this period, with a typical event lasting one or two days. These durations are longer than the likely cooling times of the \gray emitting particles, implying we see continuous acceleration rather than the result of an impulsive event such as the ejection of a single plasmoid that then cools as it propagates outwards.
Cross-correlating the \grid light curve with both X-ray and radio monitoring data, we find that the main events of \gray activity were detected while the system was in soft spectral X-ray states (\textit{RXTE}/ASM (\textit{Rossi X-ray Timing Explorer}/All-Sky Monitor) count rate in the 3-5 keV band $\gtrsim 3~\mathrm{counts~s^{-1}}$), that coincide with local and often sharp minima of the hard X-ray flux (\textit{Swift}/BAT (Burst Alert Telescope) count rate $\lesssim 0.02~\mathrm{counts}$ $\mathrm{cm^{-2}~s^{-1}}$), a few days before intense radio outbursts. This repetitive temporal coincidence between the \gray transient emission and spectral state changes of the source turns out to be the \textit{spectral signature} of \gray activity from this microquasar. These \gray events may thus reflect a sharp transition in the structure of the accretion disk and its corona, which leads to a rebirth of the microquasar jet and subsequent enhanced activity in the radio band.
The \gray differential spectrum of \mbox{Cygnus X-3} (100 MeV -- 3 GeV), which was obtained by averaging the data collected by the \grid during the \gray events, is consistent with a power law of photon index $\alpha=2.0~\pm~0.2$.
Finally, we examine leptonic and hadronic emission models for the \gray events and find that both scenarios are valid. In the leptonic model -- based on inverse Compton scatterings of mildly relativistic electrons on soft photons from both the Wolf-Rayet companion star and the accretion disk -- the emitting particles may also contribute to the overall hard X-ray spectrum, possibly explaining the hard non-thermal power-law tail seen during special soft X-ray states in \mbox{Cygnus X-3}.
}

\keywords{stars: individual: \mbox{Cygnus X-3} -- gamma rays: general -- X-rays: binaries -- radio continuum: general -- radiation mechanisms: non-thermal -- stars: winds, outflows}

\maketitle

\section{Introduction}

\mbox{Cygnus X-3} is the brightest radio source among all known microquasars and was discovered, as an X-ray source, in 1966 \citep{giacconi_67}. It is a high-mass X-ray binary, whose companion star is a Wolf-Rayet (WR) star \citep{vankerk_92} with a strong helium stellar wind \citep{szo_zdzia_08}. The system is located at a distance of about 7-10 kpc \citep{bonnet_88,ling_09}. The orbital period is 4.8 hours, as inferred from infrared \citep{becklin_73}, X-ray \citep{parsignault_72}, and \gray \citep{abdo_09} observations. Owing to its very tight orbit (orbital distance $d \approx 3 \times 10^{11}$ cm), the compact object is totally enshrouded in the wind of the companion star\footnote{The observational evidence of this strong wind can be found in the prominent attenuation of the \mbox{Cygnus X-3} power density spectrum (PDS) for frequencies above 0.1 Hz \citep{axelsson_09, koljonen_11}.}. The nature of the compact object is still uncertain\footnote{Published results suggest either a neutron star of 1.4 $M_{\odot}$ \citep{stark_03} or a black hole with a mass $\lesssim 10~M_{\odot}$ \citep{hanson_00,shrader_10}.} \citep{vilhu_09}, although a black hole scenario is favored \citep{szo_zdzia_08, szostek_08}.
In the radio band, the system shows strong flares (``\textit{major radio flares}'') reaching up to few tens of Jy. Radio observations at milliarcsec scales confirm emissions (at cm wavelengths) from both a core and a one-sided relativistic jet ($v \sim 0.81c$), with an inclination to the line-of-sight of $\lesssim14^{\circ}$ \citep{mioduszewski_01}. The radiation from the jet dominates the radio emission from the core during (and soon after) the  major flares \citep{tudose_10}.\\
\mbox{Cygnus X-3} exhibits a clear, repetitive pattern of (anti)correlations between radio and X-ray emission, and an overall anticorrelation between soft and hard X-ray fluxes \citep{mccollough_99,szostek_08}.
The most important pattern of correlations found by \citet{szostek_08} is related to the connection between radio (8.3 GHz band, GBI) and soft X-ray emissions (3-5 keV band of the \textit{Rossi X-ray Timing Explorer}/All-Sky Monitor (\textit{RXTE}/ASM)).
When the soft X-ray flux is above the transition level (3 counts/s), the source can be found in different states, depending on the level of the radio flux density. In particular, the \textit{quenched state} is characterized by a radio flux density $\leqslant$ 30 mJy and followed by a \textit{major-flaring state} with values of radio flux density $\geqslant$ 1 Jy.
It is very important to emphasize that all major radio flares have been observed after a quenched state, and in almost all cases the quenched state is followed by a major flare. 
After a major flare, a ``\textit{hysteresis}'' in the radio/soft-X-ray plane is found, because the decline in the radio flux density never occurs by means of a quenched state.

Firm detections of high-energy \grays (HE \grays: $>$100 MeV) from \mbox{Cygnus X-3}\footnote{\gray detections of \mbox{Cygnus X-3} were reported in both the 1970s and 1980s at TeV \citep{vladimirsky_73,danaher_81,lamb_82} and PeV energies \citep{samorski_83,bhat_86}. However, subsequent observations by more sensitive ground-based telescopes did not confirm TeV and PeV emission from this source \citep{oflaherty_92}. Furthermore, the \textit{COS-B} satellite could not find any clear emission from \mbox{Cygnus X-3} at MeV-GeV energies \citep{hermsen_87}, and both \textit{CGRO}/EGRET observations of the Cygnus region (1991-1994) and the first-year analysis of AGILE observations could not demonstrate that there was a solid association with the microquasar, although they confirmed a \gray detection above 100 MeV in a region including \mbox{Cygnus X-3} \citep{mori_97, pittori_09}.} were published at the end of 2009: the AGILE (Astro-rivelatore Gamma a Immagini LEggero) team found evidence that strong \gray transient emission above 100 MeV coincided with special X-ray/radio spectral states \citep{tavani_09a}, and the \lat (Large Area Telescope) collaboration announced the detection of \gray orbital modulation \citep{abdo_09}.
The peak \gray isotropic luminosity detected above 100 MeV is $L_{\gamma}\sim10^{36}~\mathrm{erg~s^{-1}}$ (for a distance of 7-10 kpc). 
{The \gray emission is most likely associated with a relativistic jet \citep{tavani_09a, abdo_09, dubus_10, cerutti_11, zdziarski_12}, but the radiative process (leptonic or hadronic) is uncertain}.

A possible leptonic scenario for \gray emission in \mbox{Cygnus X-3} was proposed by \citealp{dubus_10}: stellar ultraviolet (UV) photons are Compton upscattered to HE by relativistic electrons accelerated in the jet. The particle acceleration could take place in a shock where the jet interacts with the dense stellar wind of the WR star. The emerging picture is that of a jet with moderate bulk relativistic speed and oriented not too far from the line-of-sight.\\
The \gray modulation -- coherent with the orbital period -- suggests that the emitting region is located at distances of between $\sim$$10^{10}$ cm and $\sim$$3 \times 10^{12}$ cm ($10d$) from the compact object \citep{dubus_10,cerutti_11}. The lack of modulation at radio wavelengths and the delay ($\sim$5 days, \citealp{abdo_09}) between the onset of \gray activity and the radio flare suggest that different emission regions are linked by the collimated jet. The \gray emission -- related to inverse Compton (IC) scatterings -- most likely occurs close to the compact object, while the radio emission -- assumed to be synchrotron in origin -- occurs farther out in the jet, at an angular distance from the core of a few tens of milli-arcseconds (e.g., \citealp{tudose_07, tudose_10}), corresponding to $\sim10^{15}$--$10^{16}$ cm. The \gray modulation is due to the anisotropic efficiency of the IC scattering \citep{aharonian_81}. Thus, the \gray maximum occurs at the superior conjuction (where the compact object is behind the WR star), when relativistic electrons of the jet, moving towards the Earth, have head-on collisions with stellar UV photons. This orbital phase corresponds to the minimum of the X-ray modulation, produced in turn by the maximum of absorption/scattering by the companion's wind \citep{abdo_09,dubus_10,zdziarski_12}.

A hadronic scenario accounting for \gray emission in microquasars was discussed by \citet{romero_03,romero_05}. Their model is based on the interaction of a mildly relativistic jet with the dense wind of the companion star, and the \gray emission is due to the decay of neutral pions ($\pi^{0}$) produced by $pp$ collisions.

Furthermore, TeV emission from relativistic jet in microquasars has been predicted by several models (e.g., see \citealp{atoyan_99}). A search for very-high-energy (VHE) \grays from the microquasar GRS 1915+105 with H.E.S.S. (High Energy Stereoscopic System) was carried out, but no significant detection was found in the direction of the source \citep{hess_09}. On the other hand, hints of VHE \grays were found in Cygnus X-1 \citep{albert_07}. The Major Atmospheric Gamma-ray Imaging Cherenkov Telescope (MAGIC) observed \mbox{Cygnus X-3} several times between March 2006 and August 2009, during both its hard and soft states\footnote{The MAGIC telescope was also pointed at \mbox{Cygnus X-3} after two \gray alerts from the \grid team (the first one after the \gray event of 16-17 April 2008, and the second after the event of 13-14 July 2009, see Appendix \ref{agile_data}). In both cases, they found a $2\sigma$ upper limit, for energies above 250 GeV, of $\sim$$10^{-11}$ \flx \citep{aleksic_10}.}, but no evidence of clear VHE \gray emission from the microquasar was found: an overall $2\sigma$ upper limit to the integral flux was set at $2.2 \times 10^{-12}$ \flx for energies above 250 GeV \citep{aleksic_10}.

Here we present a comprehensive and homogeneous analysis of \mbox{Cygnus X-3} that takes into account \gray events found in the data between 2007 November 2 and 2009 July 29, during the AGILE ``pointing'' mode data-taking. We analyzed a dataset previously published by \citet{tavani_09a} and \citet{bulgarelli_12a}. We report here the whole \grid monitoring of \mbox{Cygnus X-3} during the ``pointing'' mode, to confirm that the \gray activity coincides with the same repetitive pattern of multiwavelength emission and to analyze in depth the overall \gray spectrum by assuming both leptonic and hadronic scenarios.

\section{Observations}

The AGILE scientific instrument \citep{tavani_09b} is very compact and characterized by two co-aligned imaging detectors operating in the energy ranges 30 MeV--30 GeV (GRID: \citealp{barbiellini_02, prest_03}) and 18--60 keV (Super-AGILE: \citealp{feroci_07}), as well as by both an anticoincidence system \citep{perotti_06} and a calorimeter \citep{labanti_06}. The performance of AGILE is characterized by large fields of view (2.5 and 1 sr for the \gray and hard X-ray bands, respectively) and optimal angular resolution (PSF = $3.0^{\circ}$ at 100 MeV, and PSF = $1.5^{\circ}$ at 400 MeV; see \citealp{cattaneo_11}).\\
Until mid-October 2009 AGILE had operated in ``pointing'' mode with fixed attitude; in November 2009, AGILE entered ``scanning mode'', which is characterized by a controlled rotation of the pointing axis.\\
During the ``pointing'' mode data-taking ($\sim$2.5 years), the AGILE satellite performed $\sim$100 pointings with variable exposure times (of typically 3--30 days), drifting about 1 degree per day from the initial boresight direction to match the solar-panel illumination constraints\footnote{A detailed schedule of the AGILE observations -- with the pointing starting coordinates and observation starting and ending times --  is available online at \htmladdnormallink{http://agile.asdc.asi.it/current\_pointing.html}{http://agile.asdc.asi.it/current_pointing.html}}. In this configuration, the \grid was characterized by enhanced performances in the monitoring capability of a given source, especially in the energy band 100--400 MeV (see \citealp{bulgarelli_12a} for details). Owing to the different pointing strategies of the AGILE and Fermi satellites, the high on-source cumulative exposure (between 100 and 400 MeV) of the \grid may be fundamental in the observation of this particular source.

In this paper, we report an analysis based on the \grid data collected between 2007 November 2 and 2009 July 29 (the same dataset reported by \citealp{chen_piano_11}). During this period, AGILE repeatedly pointed at the Cygnus region for a total of $\sim$275 days, corresponding to a net exposure time of $\sim$11 Ms. The detailed analysis of the dataset is presented in Appendix \ref{agile_data}. In this paper, we report seven \gray main events.

\section{The gamma-ray activity in a multiwavelength context}

\subsection{General characteristics of the gamma-ray events} \label{gen_char}

Figure~\ref{cyg_x3_all_flares_mw} shows the comprehensive multiwavelength light curve of \mbox{Cygnus X-3}, to help us analyze the pattern of multi-frequency emission. The \gray activity detected by the \grid is presented along with the hard X-ray fluxes from \textit{Swift}/BAT\footnote{\textit{Swift}/BAT (Burst Alert Telescope) transient monitor results provided by the \textit{Swift}/BAT team.} (15-50 keV), soft X-ray fluxes from \textit{RXTE}/ASM\footnote{\textit{Rossi X-ray Timing Explorer} (RXTE), All-Sky Monitor (ASM). Quick-look results provided by the \textit{RXTE}/ASM team.} (3-5 keV), and radio flux density (when available) from the AMI-LA\footnote{Courtesy of the \textit{Arcminute Microkelvin Imager} (AMI) team.} (15 GHz) and RATAN-600\footnote{Courtesy of S. Trushkin and the RATAN-600 team.} (2.15, 4.8, 11.2 GHz) radio telescopes.

The aim of the cross-correlation between the \grid light curve and the multiwavelength emission pattern is to discuss the \gray trigger criteria and compare them with those previously published by \citealp{tavani_09a}, \citealp{bulgarelli_12a}, and \citealp{corbel_12}.

Observing the light curve in Figure~\ref{cyg_x3_all_flares_mw}, as well as the detailed zooms in Figure~\ref{cyg_x3_zoom_mw}, we can note that
\begin{itemize}
  \item there is a \textit{strong anticorrelation} between the hard X-ray and \gray emission. Every local minimum of the hard X-ray light curve is associated with \gray emission detected by the \grid (see also the weak \gray event detected on 2008 June 21, modified Julian date (MJD) = 54638.58, in the plot of Figure~\ref{cyg_x3_all_flares_mw} where $\sqrt{TS}=2.77$, photon flux = $(131 \pm 61) \times 10^{-8}$ \flx).
  Conversely, every time the \grid detects \gray activity the system exhibits a deep local minimum of the hard X-ray light curve (\textit{Swift}/BAT count rate $\lesssim 0.02~\mathrm{counts}$ $\mathrm{cm^{-2}~s^{-1}}$).
  \item Every time we detect \gray activity, \mbox{Cygnus X-3} is in a soft spectral state (\textit{RXTE}/ASM count rate $\gtrsim 3~\mathrm{counts~s^{-1}}$, i.e., the transitional level defined by \citealp{szostek_08}).
  \item Every time we detect \gray episodes (Table \ref{cyg_x3_all_flares} and red points in the \grid light curve in Figure~\ref{cyg_x3_all_flares_mw}), the system is moving towards either a major radio flare (radio flux density $\gtrsim1$ Jy) or a quenched state preceding a major radio flare.
\end{itemize}
In Table \ref{cyg_x3_all_flares_states}, we report a brief synthesis of the multi-frequency pattern of emission of each main \gray event ($\sqrt{TS}\geqslant3$) detected by the \grid. The average delay between the \gray event and the subsequent radio flare is $\overline{\Delta T_2} \approx 4$ days, which is consistent with the value ($5 \pm 7$ days) found by \citet{abdo_09}.\\
If we refer to the third column in Table \ref{cyg_x3_all_flares_states}, we can see that the transient \gray emission occurs when the system is either moving into a quenched state (``pre-quenched'') or towards a radio flare (``pre-flare''), which has always been observed after a quenched state, i.e., the \gray emission is detected when the system is moving into or out of a quenched state. Hence, from a purely phenomenological point of view, the \textit{quenched state} seems to be a ``key'' condition  for the \gray emission.

We stress in general that \gray events -- always during soft states -- occur in the proximity of spectral X-ray transitions. In particular, we found that intense transient \gray emission is detected both immediately after hard-to-soft (e.g., the events of 2-3 November 2008 [MJD = 54773.17] and 20-21 June 2009 [MJD = 55003.37]) and before soft-to-hard spectral X-ray transitions (e.g., the events of 16-17 April 2008 [MJD = 54573.08], 11-12 December 2008 [MJD = 54812.39], and 13-14 July 2009 [MJD = 55025.55]). Observing Figure~\ref{cyg_x3_all_flares_mw} and the zooms in Figure~\ref{cyg_x3_zoom_mw}, we can note that strong transient \gray emission generally occurs when the system has just entered into or is moving out of a prominent minimum of the \textit{Swift}/BAT light curve (e.g., November-December 2008 [MJD $\simeq$ 54770-54815] and June-July 2009 [MJD $\simeq$ 55000-55040] events).

This comprehensive study confirms that the \gray emission conditions of \mbox{Cygnus X-3}, during the whole ``pointing'' monitoring by the AGILE satellite, agree completely with the ones found by \citealp{tavani_09a}, \citealp{bulgarelli_12a}, and \citealp{corbel_12}.

\begin{table}
\begin{center}
\caption{Main events of \gray activity ($\sqrt{TS}\geqslant3$) detected by the \grid. \textit{Column 1}: Date of the \gray event (average in MJD). \textit{Column 2}: X-ray spectral state. \textit{Column 3}: Radio-flux-density state at the time of the \gray activity. \textit{Column 4}: Time delay ($\Delta T_2$) in days between the \gray event and the major radio flare. \textit{Column 5}: Radio flux density of the major radio flare.} \label{cyg_x3_all_flares_states}
\renewcommand{\arraystretch}{1.5}
{\tiny
 \begin{tabular}{|c|c|c|c|c|}
\hline\hline
\multirow{2}{*}{MJD}      &         X-ray        &            Radio           &         $\Delta T_2$     &           Radio            \\
                          &         State        &            State           &           [days]         &      Flux Density          \\
\hline\hline
\multirow{2}{*}{54507.19} &     X-ray state      &    \multirow{2}{*}{(?)}    &   \multirow{2}{*}{(?)}   &     \multirow{2}{*}{(?)}   \\
                          &     trans. level     &                            &                          &                            \\                        
\hline
54573.08                  &         Soft         &         Pre-flare          &          $\sim 1$        & $\approx$16  Jy (11.2 GHz)\\
\hline                                                                   
54773.17                  &         Soft         &         Pre-quenched       &          $\sim 8$        & $\approx$1   Jy (15   GHz)\\
\hline                                                                   
54812.39                  &         Soft         &         Pre-flare          &          $\sim 6$        & $\approx$3   Jy (11.2 GHz)\\
\hline                                                                   
55003.37                  &         Soft         &         Pre-quenched       &            (?)           &          (?)               \\
\hline                                                                   
55025.55                  &         Soft         &         Pre-flare          &          $\sim 3$        & $\approx$3   Jy (11.2 GHz)\\
\hline                                                                   
\multirow{2}{*}{55034.88} &\multirow{2}{*}{Soft} & \multirow{2}{*}{Pre-flare} &\multirow{2}{*}{$\sim 1$} & $\approx$2.3 Jy (11.2 GHz)\\
                          &                      &                            &                          & $\approx$1.6 Jy (15   GHz)\\
\hline
\end{tabular}
}

\end{center}
\end{table}

\subsection{Peculiarities of the gamma-ray events}

\subsubsection{The gamma-ray event of 11-12 February 2008 (MJD = 54507.19)}

A more detailed discussion is needed for the \gray event of 11-12 February 2008 (MJD = 54507.19, see the upper-left panel in Figure~\ref{cyg_x3_zoom_mw}), which is a special event among the \grid detections.

\begin{landscape}
\thispagestyle{empty}
\begin{figure}
\begin{center}
\includegraphics[width=25.0cm]{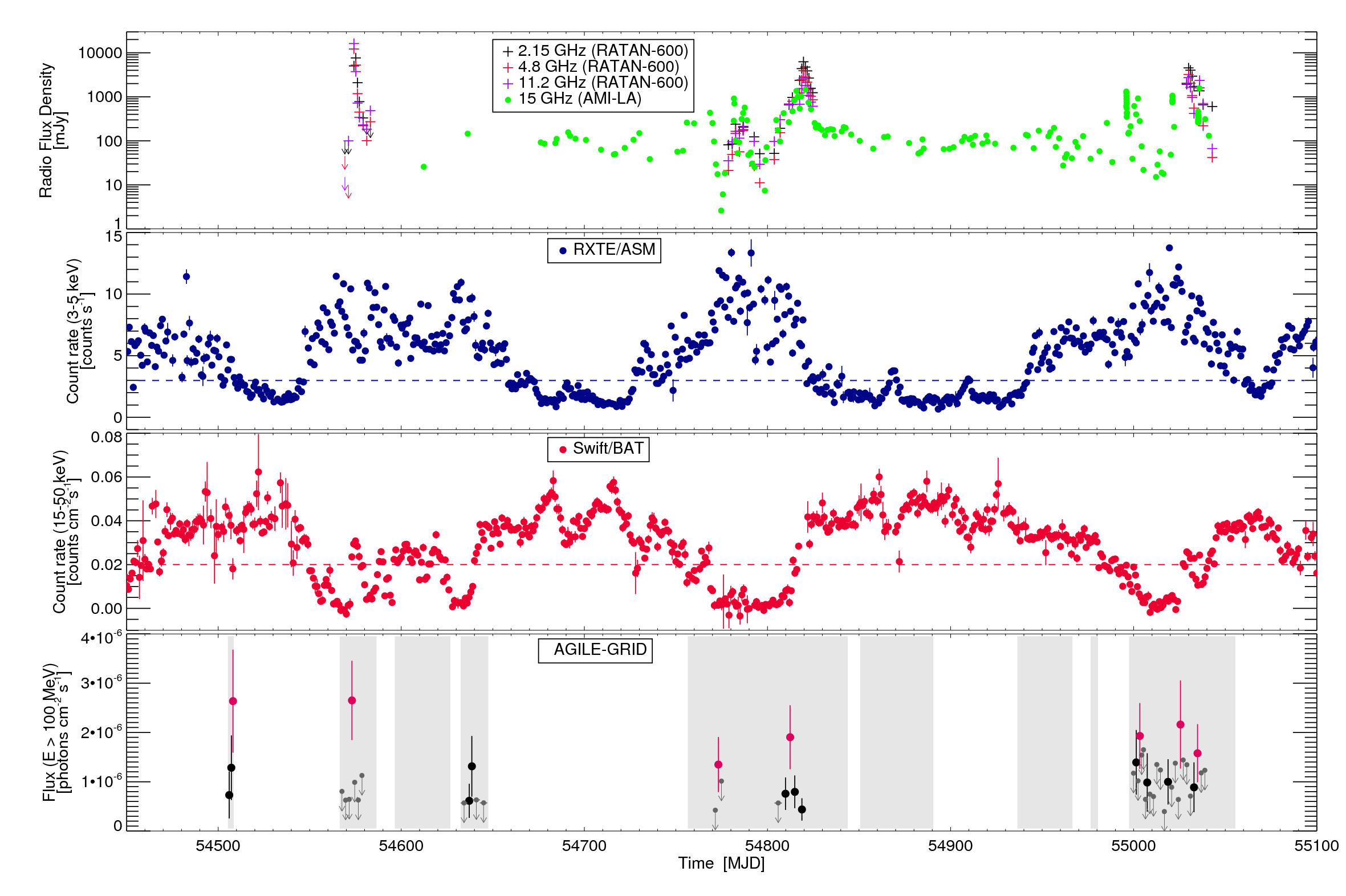}
    \caption{Multi-frequency light curve of \mbox{Cygnus X-3} from 2007 December 12 to 2009 September 26 (MJD: 54450-55100). From top to bottom: \textbf{radio} flux density [RATAN-600 (2.15, 4.8, 11.2 GHz) and AMI-LA (15 GHz)], \textbf{soft X-ray} count rate [\textit{RXTE}/ASM (3-5 keV)], \textbf{hard X-ray} count rate [\textit{Swift}/BAT (15-50 keV)], and \textbf{gamma-ray} photon fluxes [\grid (above 100 MeV)]. In the bottom panel, gray regions represent the AGILE pointing at the Cygnus region; \textit{magenta} points are the main events of \gray activity with $\sqrt{TS}\geqslant3$ (see Table \ref{cyg_x3_all_flares}), \textit{black} points are the \gray detections with $2 \leqslant \sqrt{TS} < 3$, and \textit{dark-gray} arrows are the $2\sigma$ upper limits related to $\sqrt{TS} < 2$. The dashed lines in the panels of the \textit{RXTE}/ASM and \textit{Swift}/BAT count rate represent the transition level of $3~\mathrm{counts~s^{-1}}$ and $0.02~\mathrm{counts}$ $\mathrm{cm^{-2}~s^{-1}}$ respectively (see Section~\ref{gen_char} for details).} \label{cyg_x3_all_flares_mw}
\end{center}
\end{figure}
\end{landscape}

\begin{figure*}
  \hspace*{-1.5cm}
    \begin{tabular}{cc}
\includegraphics[width=9.7cm]{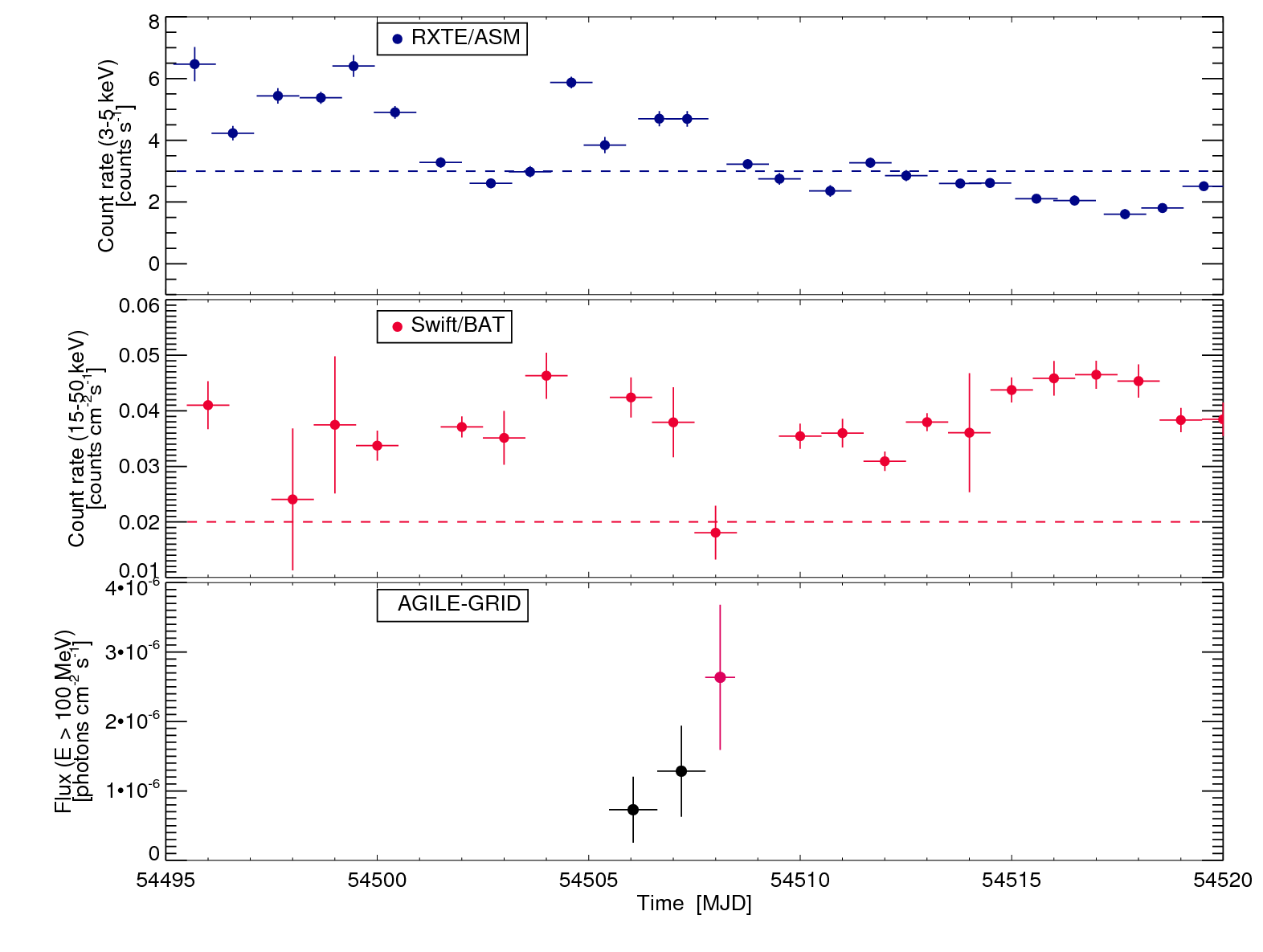}     & \includegraphics[width=9.7cm]{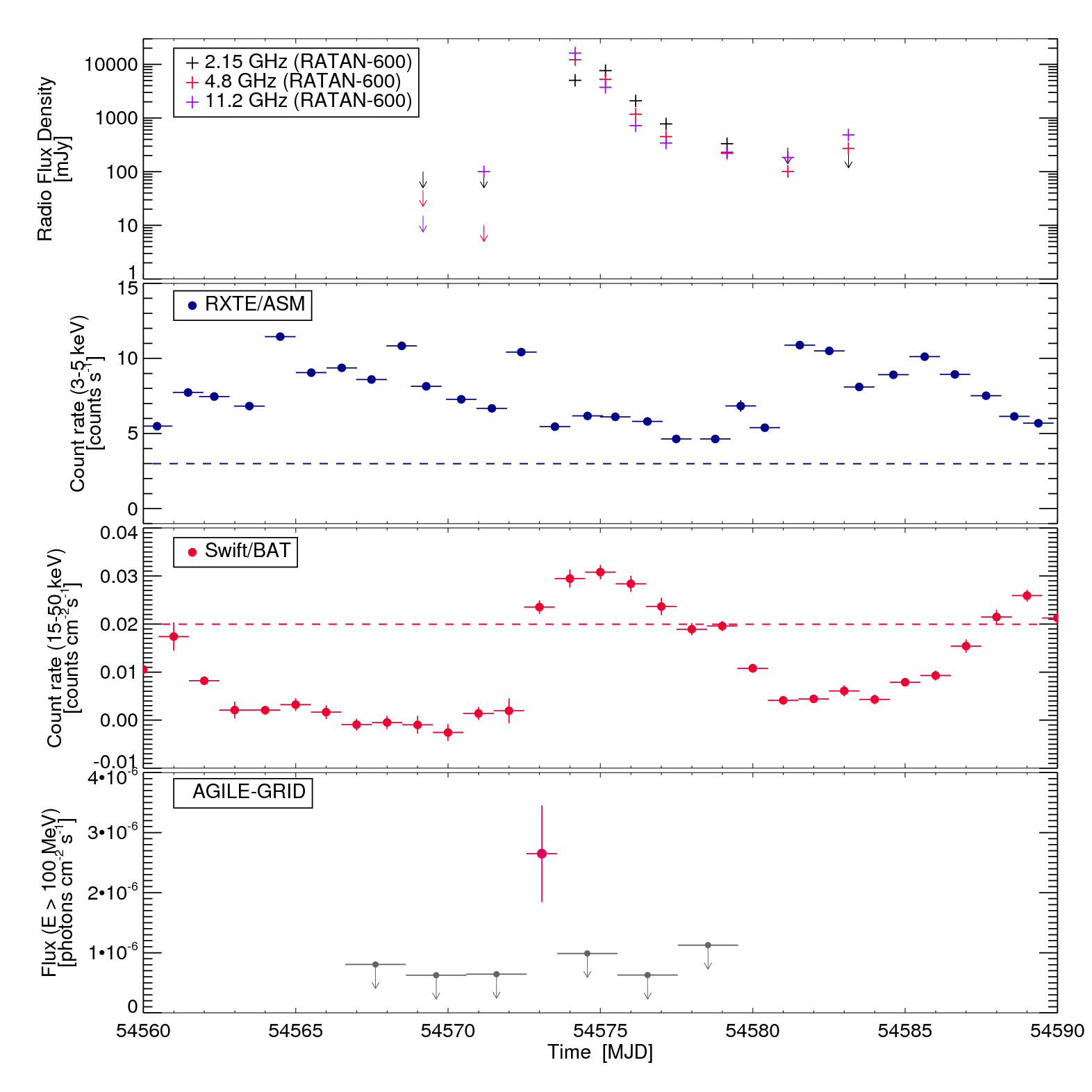}\\
\includegraphics[width=9.7cm]{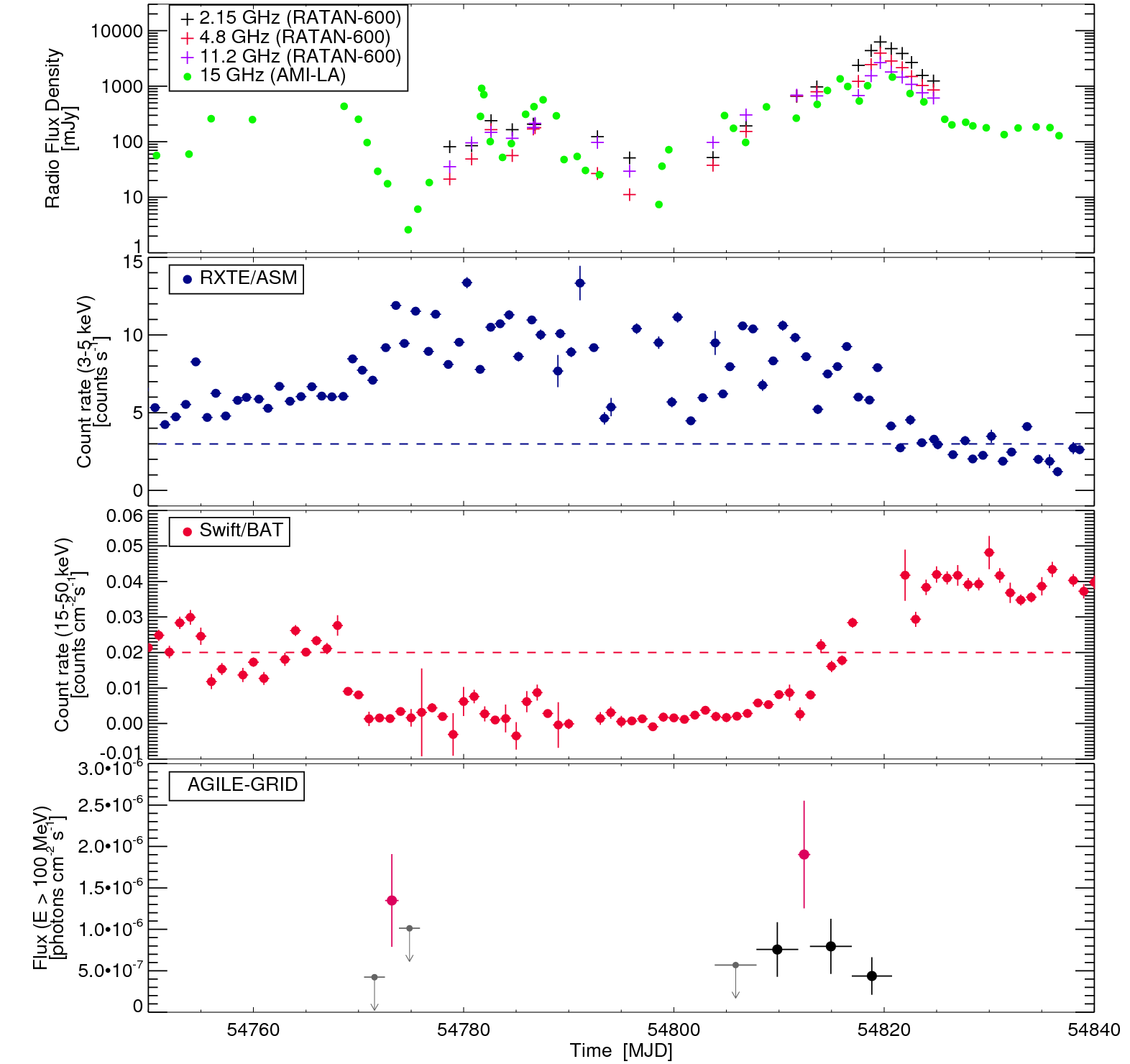} & \includegraphics[width=9.7cm]{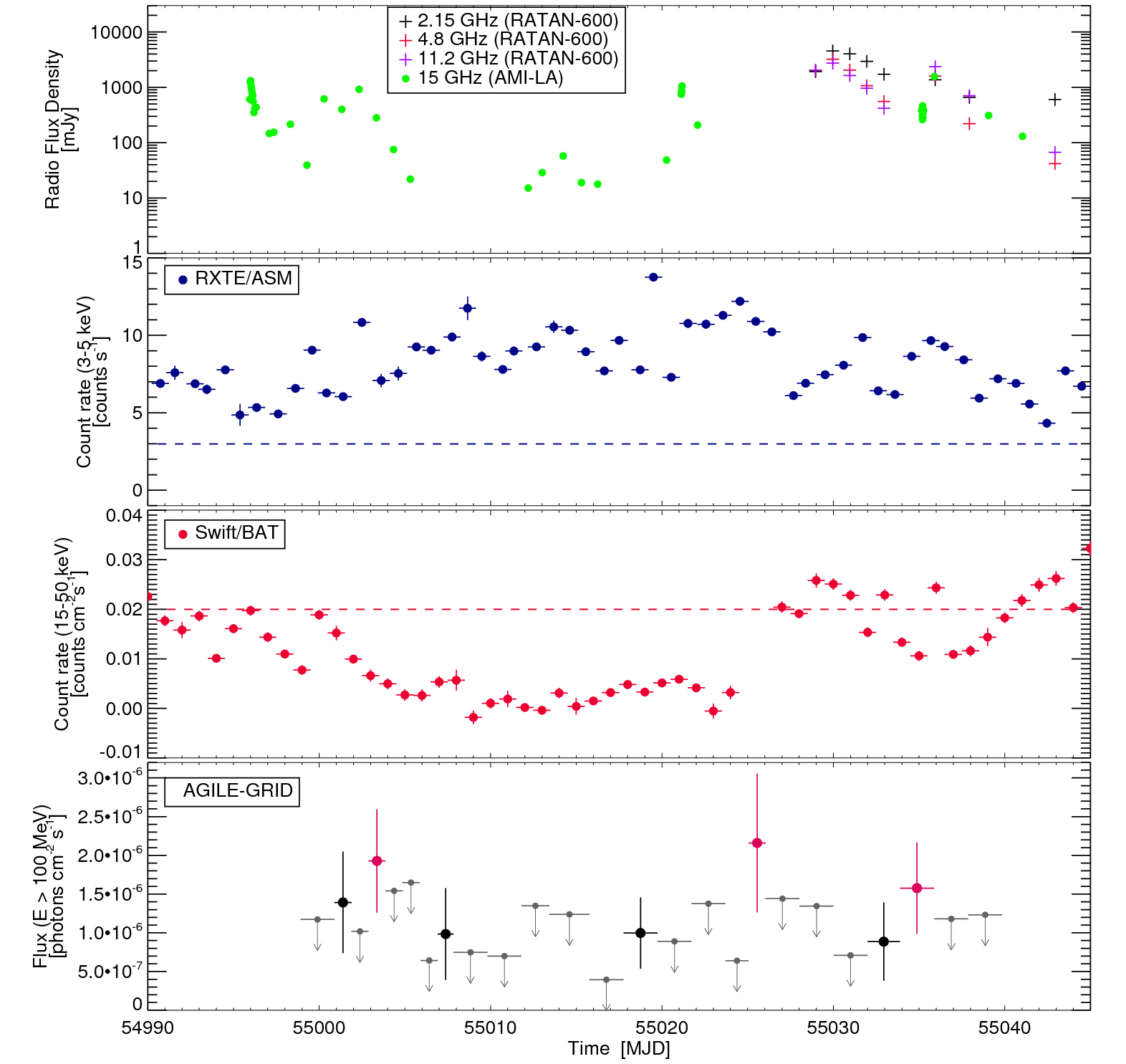}\\
    \end{tabular}
   \caption{Multi-frequency light curves centered on the main events of \gray activity detected by the \grid (detailed views of the main plot in Figure~\ref{cyg_x3_all_flares_mw}). \textit{Upper-left} plot: from 2008 January 30  to 2008 February 24 (MJD: 54495-54520). \textit{Upper-right} plot: from 2008 April 4 to 2008 May 4 (MJD: 54560-54590). \textit{Lower-left} plot: from 2008 October 11 to 2009 January 9 (MJD: 54750-54840). \textit{Lower-right} plot: from 2009 June 8 to 2009 August 2 (MJD: 54990-55045).} \label{cyg_x3_zoom_mw}
\end{figure*}

\clearpage

The \gray activity in this period occurred when \mbox{Cygnus X-3} was at the transitional level between its soft and hard X-ray spectral states (\textit{RXTE}/ASM count rate $\sim$3 $\mathrm{counts~s^{-1}}$) and coincides with a short but sharp dip in the \textit{Swift}/BAT light curve at MJD = 54508, count rate = $(0.019 \pm 0.005)$ $\mathrm{counts~cm^{-2}~s^{-1}}$, \verb+DATA_FLAG+ = 0 (data quality flag = good). Even if the event does not coincide with a \textit{bright} soft state, it seems to confirm the simultaneous \gray-event/hard-X-ray-minimum occurrences that we find in all other cases. There are no available radio data for this period.

\subsubsection{The gamma-ray event of 20-21 June 2009 (MJD = 55003.37)}

The \gray event on 20-21 June 2009 (MJD = 55003.37) occurred at the beginning of a quenched radio state (see the lower-right panel in Figure~\ref{cyg_x3_zoom_mw}). Unfortunately, we have no radio data covering the subsequent days (from the third to the ninth day after the \gray event), but we cannot exclude the presence of a major radio flare -- soon after the quenched state -- that might have escaped detection. In Figure \ref{cyg_x3_zoom_mw} (\textit{lower-right} panel), we note that the AMI-LA light curve (15 GHz) shows a subsequent radio flare $\sim$18 days after the \gray event ($\sim$13 days after the quenched state), possibly corresponding to a delayed radio burst (MJD = 55021.2, radio flux density = 1.06 Jy).

\subsubsection{The long-term gamma-ray emission June-July 2009 (MJD $\simeq$ 55000-55040)}

Finally, we note that the extended \gray activity of June-July 2009 (see the lower-right panel in Figure~\ref{cyg_x3_zoom_mw}), occurred during intense radio activity (with a high average radio flux density) coincident with a long-lasting soft X-ray spectral state. It is important to remark that this period coincides with one of the two temporal windows of strong \gray activity detected by \lat, with a peak photon flux greater than $\sim 200 \times 10^{-8}$ \flx \citep{abdo_09}. This \gray peak is simultaneous (and consistent) with the event detected by the \grid on 21-23 July 2009 (photon flux = $(158 \pm 59) \times 10^{-8}$ \flx).
These data (June-July 2009) are discussed in more detail in a dedicated paper on the \gray emission of the microquasar during the mid-2009/mid-2010 period \citep{bulgarelli_12a}.

\section{Modeling the spectral energy distribution}

By accounting for the X-ray, \gray (\grid), and TeV emission (MAGIC spectral upper limits), we modeled the multiwavelength spectral energy distribution (SED) of \mbox{Cygnus X-3} during a soft spectral state, with both the \textit{leptonic} and \textit{hadronic} scenarios. We considered an X-ray spectrum measured by \textit{RXTE}-PCA\footnote{Proportional Counter Array (PCA)} and \textit{RXTE}-HEXTE\footnote{High Energy X-ray Timing Experiment (HEXTE)} ($\sim$3--150 keV) when the source was in a ``hypersoft'' state \citep{koljonen_10}, the \grid spectrum  for the main \gray events (Figure~\ref{cyg_x3_spectrum_b20}), and the MAGIC differential flux upper limits obtained when the source was in the soft state. \citep{aleksic_10}. The hypersoft state of \mbox{Cygnus X-3}, which is a subclass of the ultrasoft state defined in \citealp{hjal_09}, is usually exhibited by the microquasar during the quenching/pre-flaring radio activity\footnote{This is an average ``hypersoft'' spectrum related to 28 pointed \textit{RXTE} observation between February 2000 and January 2006 (see the supporting information of \citealp{koljonen_10} for details).}. This X-ray spectral state is characterized by a weak and hard power-law tail ($\alpha$= 1.7--1.9) of non-thermal origin.\\
Thus, we analyzed a pattern of multiwavelength datasets that, even if not acquired simultaneously, are qualitatively consistent because they all refer to the same spectral state of \mbox{Cygnus X-3}: the X-ray and \grid datasets are related to the soft-state activity preceding the radio major flares and the TeV data are related to the soft-state activity following the radio major flares (MAGIC has never observed \mbox{Cygnus X-3} during its pre-flaring radio states).

\subsection{A leptonic scenario}

We modeled the multi-frequency SED by assuming a simple leptonic scenario in which a plasmoid of high energy electrons/positrons, injected into the jet structure, upscatters via inverse Compton interactions soft seed photons from both the WR star and the accretion disk. 

Our aim is to analyze a possible link between the power-law tail of this special soft X-ray spectral state and the \gray emission detected by the \grid.

The physical parameters of the photon field are literature-based. We modeled the X-ray data with a black body (BB) spectrum\footnote{Our modeling is a simplification: we assume that the bump in the X-ray emission during the ``hypersoft'' state can be modeled with a simple BB component, which is a very good approximation for our purposes. In this state, the overall X-ray emission is totally dominated by strong BB emission from the accretion disk. Nevertheless, more accurate modeling should be based on a Comptonized BB spectrum of the corona (see \citealp{koljonen_10} for details).} characterized by a temperature $T_{bb}\approx1.3$ keV, which is consistent with the typical characteristic temperature of the disk during the hypersoft/ultrasoft state \citep{hjal_09,koljonen_10}, and a $L_{bb}\approx 8 \times 10^{37}~\mathrm{erg~s^{-1}}$. The main parameters that we used for the WR star are $T_{\star}=10^5~\mathrm{K}$ and $L_{\star} \approx 10^{39}~\mathrm{erg~s^{-1}}$ (see \citealp{dubus_10}). The WR star is assumed to emit UV photons isotropically. We modeled the average \gray emission in the orbital phase. Thus, the WR photons are assumed to come mainly from the side of the jet and collide with the relativistic leptons via IC scattering processes.

We carried out two different models: in the first one (leptonic model \textit{``A''}), the plasmoid interacts with the soft photon bath ``close'' to the disk (the star-plasmoid distance is $R \approx d \approx 3 \times 10^{11}$ cm), whereas in the second one (leptonic model \textit{``B''}) the interaction region is ``far away'' from the accretion disk (the star-plasmoid distance is $R \approx 10d \approx 3 \times 10^{12}$ cm).\\
For both models, the inclination of the jet to the line of sight is assumed to be $i = 14^{\circ}$, and the plasmoid is assumed to be spherical (radius $r = 3 \times 10^{10}$ cm) with a bulk motion characterized by a Lorentz factor of $\Gamma=1.5$ ($v = \sqrt{5}
c/3$). The population of electrons is modeled by a broken-power-law spectral distribution, with spectral indices $\alpha_1=2.2$, $\alpha_2=4.0$, $\gamma_{min}=1$, $\gamma_{max}=10^5$, and an energy break of $\gamma_b=4\times 10^3$
\begin{equation} \label{broken_plaw}
 \frac{\de N}{\de \gamma \de \mathrm{V}}=\frac{K_e \, \gamma_b^{-1}}{\left(\frac{\gamma}{\gamma_b}\right)^{\alpha_1} + \left(\frac{\gamma}{\gamma_b}\right)^{\alpha_2}} ~~~~~~~~~~~[\gamma_{min} \leqslant \gamma \leqslant \gamma_{max}] ~~.
 \end{equation}
The spectral indices and the energy break of the electron distribution are the best-fit values for the \grid spectral shape. The distribution of electrons/positrons is assumed to be isotropic in the plasmoid rest frame (the jet comoving frame).
We adopted the Klein-Nishina formula to describe the Compton scattering of soft photons by a cloud of mildly relativistic leptons \citep{aharonian_81}.

\begin{figure}[!tbp]
 \begin{center}
    \includegraphics[width=9cm]{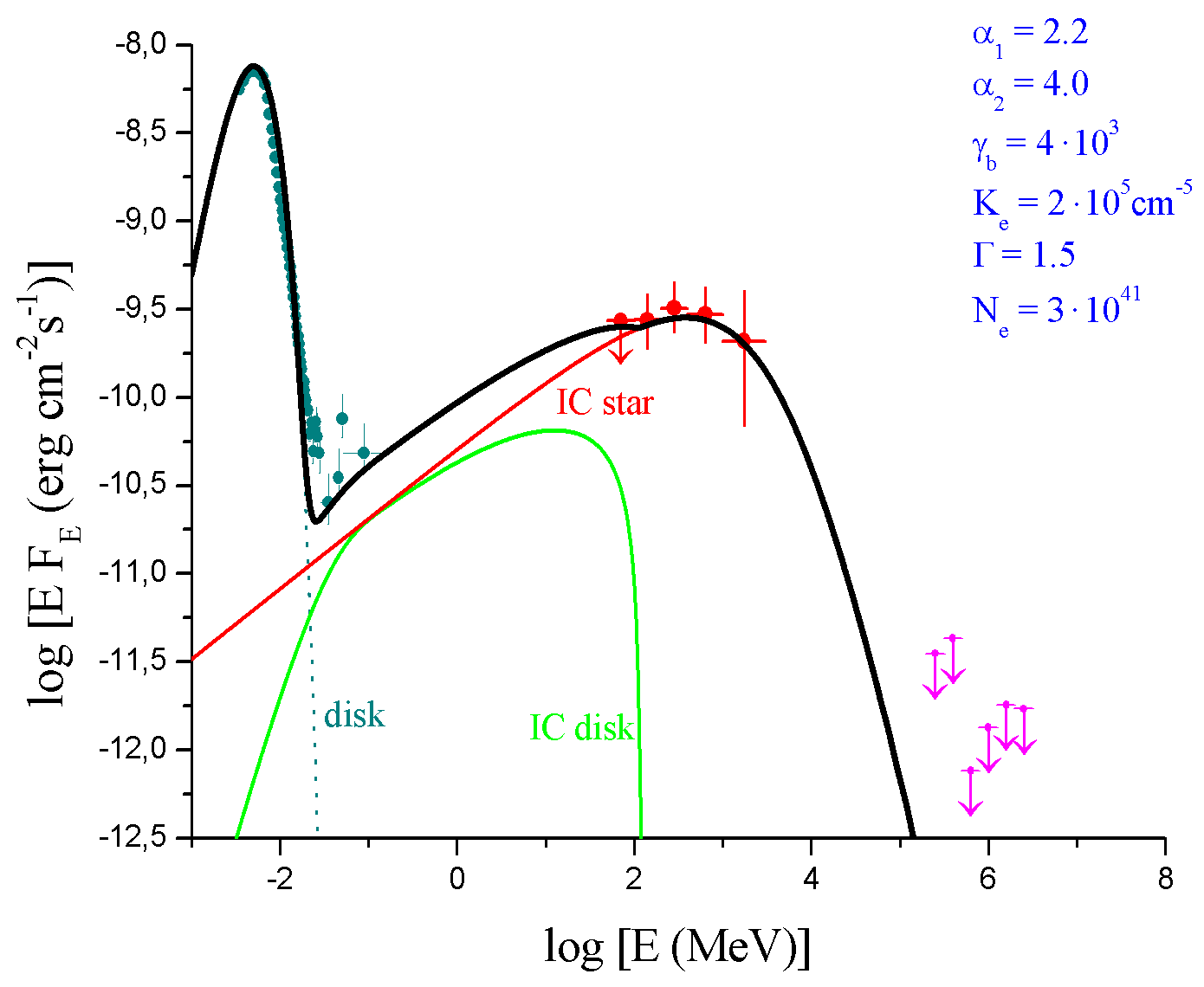}
    \caption{Multiwavelength SED of \mbox{Cygnus X-3} during the main \gray events (non-simultaneous data) and the leptonic model \textit{``A''} (see main text). \textit{Blue circles}: X-ray average ``hypersoft'' spectrum \citep{koljonen_10}, \textit{RXTE}-PCA and \textit{RXTE}-HEXTE data ($\sim$3 to $\sim$150 keV); \textit{red circles}: \grid energy spectrum (50 MeV to 3 GeV) of the main \gray episodes (Figure~\ref{cyg_x3_spectrum_b20} and \ref{cyg_x3_agile_fermi_only}); \textit{magenta arrows}: MAGIC differential flux upper limits (95\% C.L.), 199--3155 GeV, related to soft spectral state \citep{aleksic_10}. Spectral components of the model are the BB emission from the disk (blue short-dashed line), IC scattering of the soft photons from the accretion disk (green solid line), and IC scattering of the soft stellar photons (red solid line). The global SED model curve is indicated by a black solid line.} \label{cyg_x3_mw_spectrum_lep_a}
\end{center}
\end{figure}

In the leptonic model \textit{``A''}, the distance from the star to the plasmoid location is assumed to be $R \approx 3 \times 10^{11}$ cm ($R \approx d$), i.e., the plasmoid in the jet is very close both to the compact object and the accretion disk. The distance between the plasmoid center and the compact object is $H \approx 3 \times 10^{10}$ cm, i.e., $H \approx r$. The results of this modeling are presented in Figure~\ref{cyg_x3_mw_spectrum_lep_a}. The electron number density of the plasmoid is $n_e \approx 3 \times 10^9$ electrons $\mathrm{cm^{-3}}$ (the prefactor in Eq.~(\ref{broken_plaw}) is $K_e = 2 \times 10^5$ $\mathrm{cm^{-3}}$, and the integrated number of electrons is $N_e = 3 \times 10^{41}$).
We took into account the $\gamma \gamma$ absorption (for $e^{\pm}$ pair production) of the IC \gray photons by the X-ray photons from the accretion disk. We assumed that the distribution of the disk photons is fully isotropized by the stellar wind in the observer frame. This implies that the \gray photosphere (i.e., where $\tau_{\gamma \gamma}\geqslant1$) has a radius of $\sim10^{10}$ cm \citep{cerutti_11}. With these assumptions, the lowest part of the plasmoid is within the \gray photosphere\footnote{Assuming that the photosphere radius is equal to the plasmoid radius, i.e., $3 \times 10^{10}$ cm, the fraction of the plasmoid volume inside the \gray photosphere is $\sim$32\%.}. The spectral component related to the IC scatterings of the disk photons (green curve) is actually produced in this region, very close to the disk, where the X-ray photon density as well as the optical depth is high. Since $\tau_{\gamma \gamma} > 1$, this component displays a sharp cut-off energy at $\sim$100 MeV (i.e., the threshold for $e^{\pm}$ production, given the characteristic energies of the disk photons). On the other hand, the spectral component related to the IC scatterings of the stellar wind photons (red curve) does not show any cut-off energy, because it is mainly produced in the farthest part of the plasmoid (outside the \gray photosphere, for distances greater than $\sim10^{10}$ cm from the disk), where the $\gamma \gamma$ absorption by the X-ray disk photons is negligible. Thus, we deduced that in our geometry the plasmoid volume outside the \gray photosphere emits the bulk of the \gray emission above 100 MeV via IC processes acting on stellar photons (see Figure~\ref{cyg_x3_mw_spectrum_lep_a}).\\
In model \textit{``A''}, assuming a lepton injection rate of $\dot{N}_e = n_e~\pi r^2 v \approx 2 \times 10^{41}~\mathrm{leptons~s^{-1}}$, the jet kinetic luminosity for the leptons ($L_{kin,~e} = \dot{N}_e~\Gamma~m_{e} c^2$) would be $L_{kin,~e}^{A} \approx 2 \times 10^{35}$ $\mathrm{erg~s^{-1}}$.

\begin{figure}[!tbp]
 \begin{center}
    \includegraphics[width=9cm]{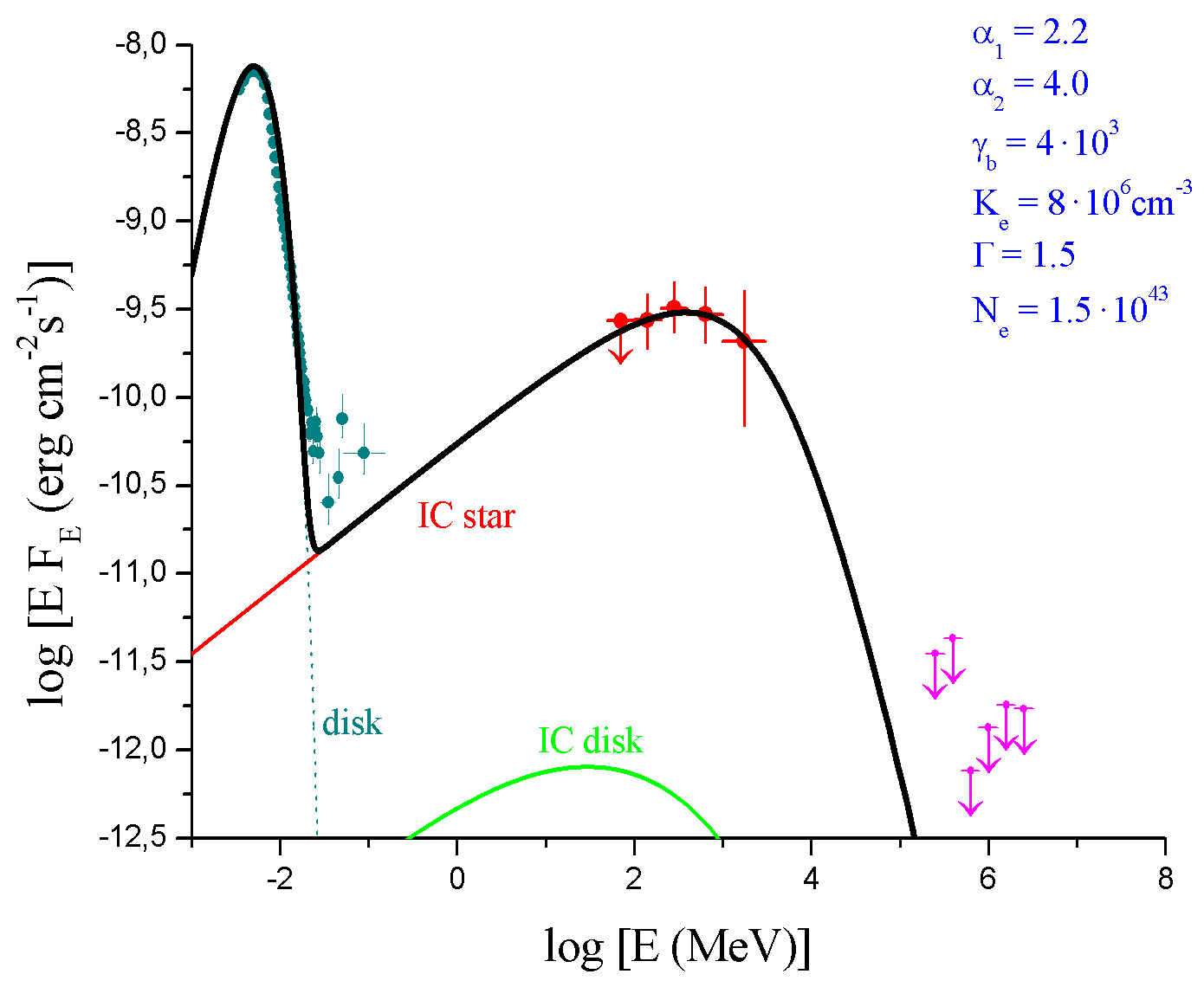}
    \caption{Multiwavelength SED of \mbox{Cygnus X-3} during the main \gray events (non-simultaneous data) and the leptonic model \textit{``B''} (see main text). Spectral components of the model are the BB emission from the disk (blue short-dashed line), IC scattering of the soft photons from the accretion disk (green solid line), and IC scattering of the soft stellar photons (red solid line). The global SED model curve is indicated by a black solid line. For a detailed description of the datasets, see caption to Figure~\ref{cyg_x3_mw_spectrum_lep_a}.} \label{cyg_x3_mw_spectrum_lep_b}
\end{center}
\end{figure}

In the leptonic model \textit{``B''}, the distance from the star to the plasmoid is assumed to be $R \approx 3 \times 10^{12}$ cm ($R \approx 10d$), i.e., the plasmoid in the jet is far away from the compact object and the accretion disk. The distance between the plasmoid center and the compact object is $H \approx 3 \times 10^{12}$ cm, i.e., $R \approx H$. We assumed that the disk photons enter the plasmoid mainly from behind. The results of this modeling are shown in Figure~\ref{cyg_x3_mw_spectrum_lep_b}. The electron density of the plasmoid is $n_e \approx 1.5 \times 10^{11}$ electrons $\mathrm{cm^{-3}}$, where the prefactor in Eq.~(\ref{broken_plaw}) is $K_e = 8 \times 10^6$ $ \mathrm{cm^{-3}}$ and the integrated number of electrons is $N_e = 1.5 \times 10^{43}$. In this model, the spectral component related to the IC scatterings of disk photons (green curve) is negligible compared to the IC component of soft photons from the star (red curve).
We note that the ``IC disk'' component does not show any cut-off energy related to the $\gamma \gamma$ absorption by X-ray photons, because the IC \grays are produced well outside the \gray photosphere (at distances $\gg 10^{10}$ cm).\\
In model \textit{``B''}, assuming a lepton injection rate of $\dot{N}_e = n_e~\pi r^2 v \approx 10^{43}~\mathrm{leptons~s^{-1}}$, the jet kinetic luminosity for the leptons ($L_{kin,~e} = \dot{N}_e~\Gamma~m_{e} c^2$) would be $L_{kin,~e}^{B} \approx 10^{37}$ $\mathrm{erg~s^{-1}}$.

In these models, the expected VHE \gray emission would be very faint. These expectations are consistent with the MAGIC upper limits, and might explain the lack of TeV bright detections during soft states.

\subsection{A hadronic scenario}

We also considered a ``hadronic scenario'' for \gray production
from \mbox{Cygnus X-3}. In our model, we used the same formalism adopted by \citet{romero_03}.
In this case, the compact source is assumed to
eject a flux of mildly relativistic hadrons (mostly protons) at
the base of the jet. These protons are first accelerated near the
compact object and then propagate along the jet interacting with
the gaseous surroundings provided by the WR companion mass-outflow.
The resulting proton-proton ($pp$) collisions can copiously
produce pions and \grays resulting from neutral pion decays.

The proton distribution in the jet is assumed to be isotropic in the jet comoving frame,
with an energy spectrum described by a power law with a high energy cut-off
\begin{equation} \label{plaw_coff}
 \frac{\de N}{\de \gamma \de \mathrm{V}}=K_p \, \gamma^{-\alpha}~{\rm exp}(-\gamma/\gamma_c) ~~~~~~~~~~~[\gamma \geqslant \gamma_{min}]
\end{equation}
with $\alpha=3$, $\gamma_{min}=1$, and $\gamma_c=100$.
The spectral index of the distribution is the best-fit value for the \grid spectral shape. 
We set the energy cut-off value at $\gamma_c=100$ so that the total SED is consistent with the spectral constraints of the MAGIC upper limits.

The ejected protons interact with the hadronic
matter of the WR strong wind. The inelastic hadronic scatterings produce neutral
pions that subsequently decay into \grays.
We adopted the same formula for the cross section $\sigma_{pp}(\gamma)$ of inelastic $pp$ interaction
reported by \citet{kelner_06}.
We assumed that the injected protons in the jet interact with the gas of the wind along a
cylindrical column of matter\footnote{A cylindrical configuration of the jet corresponds to setting $\epsilon=0$
in the formalism adopted by \citet{romero_03} to describe the jet radius dependence on the axis,
$r(z)= \xi z^{\epsilon}$.} with a
radius $r = 3 \times 10^{10}$ cm and a height of $H \approx 3 \times
10^{12}$ cm (this height provides the interesting part of the
cylinder in which most of the interactions take place). 
In analogy with the leptonic models, we assumed for the jet a bulk Lorentz factor of $\Gamma=1.5$, 
and an inclination to the line of sight of $i = 14^{\circ}$.
To quantify the density of matter in the WR wind, we assumed that
the companion star has a mass-loss rate of $\dot{M} \sim 10^{-5}
M_{\odot}~\mathrm{yr}^{-1}$ and the speed of the wind is $v_{wind}
\sim 1000~\mathrm{km~s^{-1}}$ \citep{szo_zdzia_08}. By integrating the
density of matter in this cylinder expressed in terms of the number
density of protons ($\rho \sim 1/R^2$, where $R$ is the distance
from the star), we find that the total number of protons from the
wind in this column is $N_{p,wind} \approx 3.7 \times 10^{45}$.

\begin{figure}[!tbp]
 \begin{center}
    \includegraphics[width=9cm]{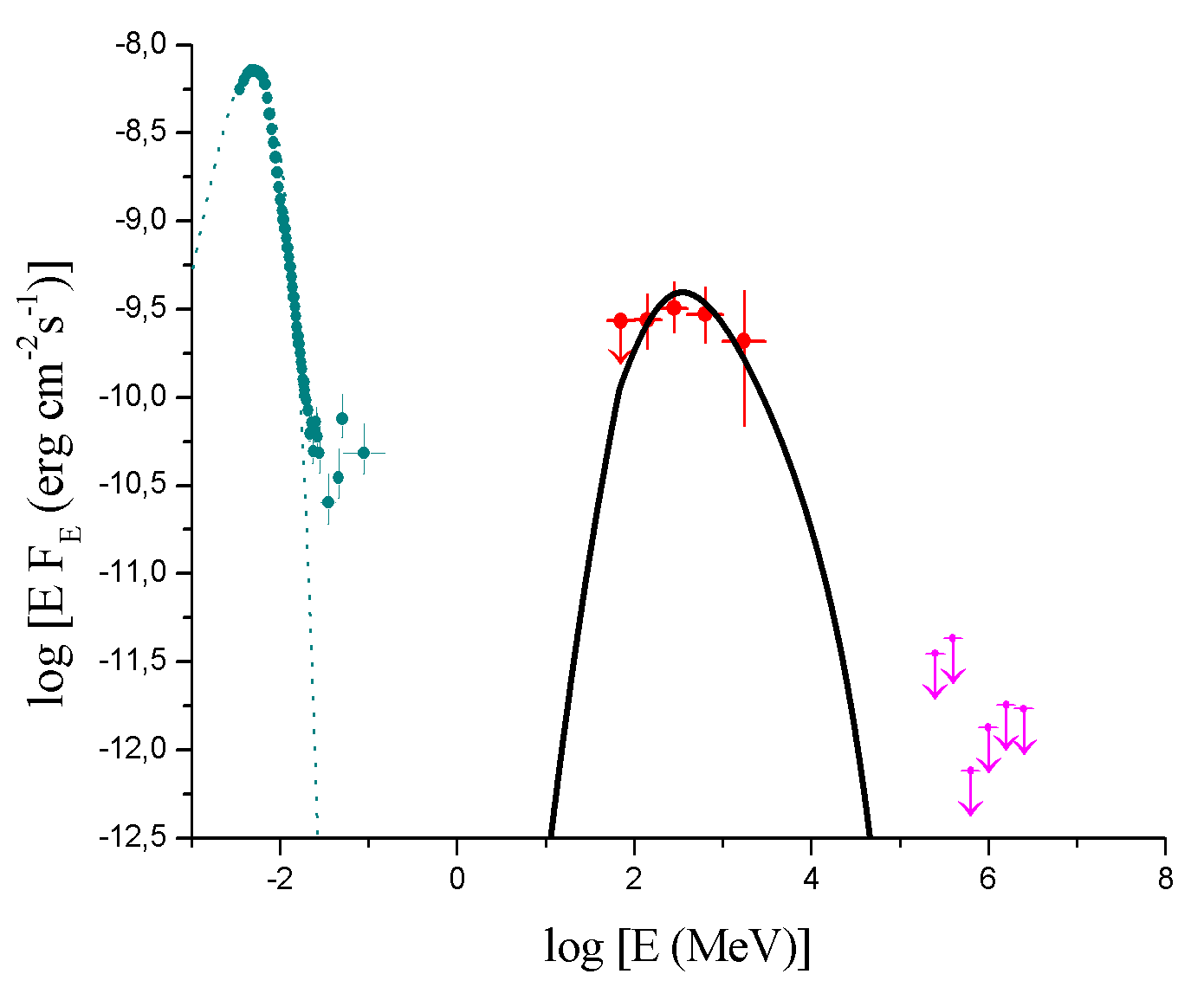}
    \caption{Multiwavelength SED of \mbox{Cygnus X-3} during \gray events (non-simultaneous data) and a hadronic model (see main text). Black body emission from the disk (blue short-dashed line), and \gray emission from $\pi^0$-decays (black solid line) are plotted. For a detailed description of the datasets, see caption to Figure~\ref{cyg_x3_mw_spectrum_lep_a}.} \label{cyg_x3_mw_spectrum_had}
\end{center}
\end{figure}

We considered a variety of proton injection rates in the jet, $\dot{N}_{p,jet}$.
The result of our best-fit hadronic model for \mbox{Cygnus X-3} is reported
in Figure~\ref{cyg_x3_mw_spectrum_had}. In this case, the integrated number
of protons injected in the jet is $N_{p,jet} \approx 9.0 \times 10^{42}$,
and the average proton number density in the column of interest is
$n_{p,jet} \approx 1.1 \times 10^{9}$ protons $\mathrm{cm^{-3}}$.
The proton injection flux in the jet of our best-fit model
turns out to be $\phi_{p,jet} \approx 2.4 \times 10^{19}$ protons
$\mathrm{cm^{-2}~s^{-1}}$, which corresponds to a proton injection rate of
$\dot{N}_{p,jet} \approx 6.7 \times 10^{40}$ protons $\mathrm{s^{-1}}$.

In \mbox{Cygnus X-3}, the corresponding jet kinetic luminosity for the hadrons ($L_{kin,~p} = \dot{N}_p~\Gamma~m_{p} c^2$) would be $L_{kin,~p} \approx 1.5 \times 10^{38}$ $\mathrm{erg~s^{-1}}$. This value is consistent with the average bolometric luminosity of the hypersoft state, $L^{HYS}_{bol} \approx 1.2 \times 10^{38}$ $\mathrm{erg~s^{-1}}$ \citep{koljonen_10}.
Moreover, $L_{kin,~p}$ is lower than the Eddington accretion limit for the system, which is $L_{Edd} \approx 10^{39}$ $\mathrm{erg~s^{-1}}$ assuming that the compact object is a black hole with a mass of $M_x \approx 10 M_{\odot}$.

\section{Discussion}

In the context of a leptonic scenario, we find that most of the \gray emission above 100 MeV is due to IC scatterings of stellar wind photons by relativistic electrons (see Figures \ref{cyg_x3_mw_spectrum_lep_a} and \ref{cyg_x3_mw_spectrum_lep_b}), according to the results of \citealp{dubus_10} and \citealp{zdziarski_12}. We note that the IC cooling times are very short ($t_{IC}$ $\sim$1--10 s). Thus, the observed time-scale of \gray emission (1-2 days) and the strong orbital modulation ($\sim$4.8 hours) detected in \grays \citep{abdo_09} impose a continuous injection of accelerated particles into the jet. The \gray emission, as also noted by \citet{zdziarski_12}, cannot be related to a single injection of a relativistic plasmoid in the jet. Our leptonic picture may suggest that there is a substructure. We find that the innermost part of the jet, where the density of X-ray disk photons is relatively high, could contribute significantly to the hard X-rays at $\sim$100 keV. In model ``\textit{A}'', the spectral component related to IC scatterings of soft photons from the disk (green curve) gives a substantial contribution to the overall model below $\sim$100 MeV that is consistent with the hard tail of the X-ray dataset (see Figure~\ref{cyg_x3_mw_spectrum_lep_a}). This model suggests that there is a possible spectral link between the power-law tail of ultrasoft/hypersoft state and the \gray emission detected by the \grid. 
On the other hand, in the region of the jet outside the \gray photosphere related to pair production on disk photons ($H \gtrsim 10^{10}$ cm), the ``IC disk'' component becomes very low at $\sim$100 keV and the overall contribution to the hard X-rays is negligible (Figure~\ref{cyg_x3_mw_spectrum_lep_b}). Thus, according to model ``\textit{B}'', if the region of \gray emission is far away from the compact object, the power-law tail in the hard X-ray band cannot be accounted for by IC processes in the jet.
For this reason, we assumed two ``extreme'' cases in the spatial configuration of the \gray emitting blob, which is located at $H \approx 3 \times 10^{10}$ cm  (model ``\textit{A}'') and $H \approx 3 \times 10^{12}$ cm (model ``\textit{B}''). \citet{dubus_10} found that the \gray emitting region is located at a distance $H \lesssim 10 d \approx 3 \times 10^{12}$ cm. By assuming that the compact object is a black hole with a mass of $M_x = 20 M_{\odot}$, their simulation found that the best fit to the \gray modulation was for $H \approx 3 \times 10^{11}$ cm. Hence, the bulk of the \gray emission is possibly produced at an intermediate configuration between models ``\textit{A}'' and  ``\textit{B}''.\\
The leptonic picture \textit{``B''} is qualitatively consistent with the one proposed by \citet{zdziarski_12} to explain the \lat data published in \citet{abdo_09}, even if the peak energy of the IC bump is quite different: in our models the peak energy is between $\sim$100 MeV and $\sim$1 GeV, whereas in their models the peak energy is between $\sim$3 MeV and $\sim$50 MeV. This difference could be due to the unequal spectral indices of the \gray spectra detected by the \grid and \lat (see Figure~\ref{cyg_x3_agile_fermi_only}).
Moreover, \citet{zdziarski_12} assumed that the electrons are injected in the jet with a power-law rate for $\gamma_1 \leqslant \gamma \leqslant \gamma_2$. The electrons subsequently lose energy via Compton, synchrotron, and adiabatic losses and form a distribution below $\gamma_1$. Thus, they demonstrated that models with $\gamma_1 < 10^3$ contribute significantly to the hard X-rays, which appears to conflict with the observed orbital modulation at $\sim$100 kev during the \gray emitting intervals, which is out of phase with the \gray modulation at $>100$ MeV \citep{zdziarski_press}. This phase misalignment would rule out any substantial contribution of the jet to the hard X-rays, in (apparent) contradiction with our findings for model ``\textit{A}''. We first remark that in our models the energy distributions of the accelerated particles are assumed to be ``steady-state'' spectra, arising from physical cooling processes, and not injected spectra.  Our leptonic models are then based on the IC scatterings of both UV stellar photons and X-ray disk photons, whereas \citet{zdziarski_12} consider the UV stellar photons only (neglecting the X-ray photons from the accretion disk). By observing Figure~\ref{cyg_x3_mw_spectrum_lep_a}, we note that at $\sim$100 keV the contribution of the ``IC disk'' component (green curve) is equivalent to the contribution of the ``IC star'' component (red curve). However, the latter has a modulated emission (the jet-wind geometry is anisotropic during the orbital phase), whereas the former is unmodulated (the jet-disk geometry does not significantly change with orbital phase). Thus, the effective unabsorbed modulation at $\sim$100 keV -- that is a superposition of a modulated and a unmodulated component -- should have a lower amplitude than the effective modulation at energies $>100$ MeV, which is actually related only to the modulated ``IC star'' component. Moreover, anisotropic absorption effects in the wind could strongly affect the unabsorbed 100 keV emission and produce an observed modulation in phase with the soft X-ray band.\\
Finally, we note that, for energies higher than $\sim$10 GeV, \grays are above the threshold for pair production on stellar photons. Nevertheless, \citealp{zdziarski_12} demonstrated that, for a similar choice of geometrical parameters, the value of the optical depth is moderate and peaks at $\sim$0.1--1 TeV. Thus, the $\gamma \gamma$ absorption by UV stellar photons was neglected in our (leptonic and hadronic) models.
\\
In the context of a hadronic scenario, we used a model similar to the one proposed by \citet{romero_03}. The only substantial difference consists in the jet geometry: we used a cylindrical model, whereas they used a conical configuration. We found that a simply hadronic model can account for the \gray spectrum detected by the \grid, by assuming a reasonable proton injection rate in the jet. It is interesting to compare our best-fit hadronic injection rate for \mbox{Cygnus X-3} with the value deduced for the microquasar SS433, which is known to produce a quasi-steady jet of hadronic nature \citep{migliari_02}. SS443 is characterized by jet mass-ejection rates near $\dot{M}_{jet} \approx 5 \times 10^{-7} \, M_{\odot} \, yr^{-1}$ \citep{konigl_83,fabrika_87,reynoso_08}, which corresponds to a proton injection rate of $\dot{N}_{p,jet}^{~SS443} \approx 1.9 \times 10^{43}$ protons $\mathrm{s^{-1}}$. Thus, we have that  $\dot{N}_{p,jet}^{~Cyg-X-3} \approx 3.5 \times 10^{-3} ~ \dot{N}_{p,jet}^{~SS443}$. SS433 ejects hadrons in a quasi-steady fashion, whereas \mbox{Cygnus X-3} is supposed to eject hadrons in a highly variable regime with a lower injection rate.\\
Our hypothesis for a hadronic interpretation of \gray emission from \mbox{Cygnus X-3} needs to be supported by information that at the moment remains unavailable, such as hadronic emission lines in the flare spectra and a precise characterization of the \gray spectrum at energies below 100 MeV that should show the characteristic decrement of neutral pion emission. Furthermore, hadronic mechanisms, besides emitting strong \gray radiation via $\pi^0$-decay, would produce an intense flux of high-energy neutrinos, emerging from the decay of secondary charged mesons produced in $pp$ collisions. Hence, a firm simultaneous detection of strong neutrino flux and \gray activity from \mbox{Cygnus X-3} would represent the signature of a dominant hadronic mechanism in the relativistic jet.
In our hadronic scenario, owing to the temporal coincidence of the \gray/radio flares, we implicitly assume that the hadronic component of the jet provides the main contribution to the \gray emission, and the leptonic component produces -- via synchrotron emission process -- the strong radio flares far away from the compact object.  In addition to \grays from $\pi^0$-decays, hadronic \textit{pp} interactions are expected to produce a population of secondary electrons (and positrons) from the decay of charged pions ($\pi^{\pm}$). These secondary leptons can contribute to the emission in the radio band via synchrotron processes and in \grays (marginally with respect to the contribution by $\pi^0$-decay) via IC and bremsstrahlung processes.

\section{Conclusions}

Several events of \gray activity were detected by the \grid from \mbox{Cygnus X-3} while the system was in a special radio/X-ray spectral state: intense \gray activity was detected during prominent minima of the hard X-ray light curve (corresponding to strong soft X-ray emission), a few days before intense radio outbursts (major radio flares). This temporal repetitive coincidence turned out to be the spectral signature of \gray activity from this puzzling microquasar, which might open new areas to study the interplay between the accretion disk, the corona, and the formation of relativistic jets. The simultaneous strong soft X-ray emission from the disk and \gray emission from the jet preceding the intense radio outbursts are consistent with a scenario in which the hot thermal corona ``dissolves'' and the accretion power from the disk directly charges the jet, emitting \grays and, subsequently, radio outbursts (via synchrotron processes) far from the compact object. 

The \gray detections of \mbox{Cygnus X-3} provide new constraints on emission models for this powerful X-ray binary, indicating that hybrid-Comptonization mechanisms \citep{coppi_99} alone cannot account for the \gray fluxes detected by AGILE and Fermi above 100 MeV, unless we assume unrealistic physical parameters \citep{cerutti_11}. This implies that the corona cannot be the site of the \gray emission. We found that the innermost part of the jet (distances $\lesssim 10^{10}$ cm from the compact object) could provide a strong contribution to the hard X-rays at $\sim$100 keV during  the \gray emitting interval, while the farthest part (distances $\gtrsim 10^{10}$ cm from the compact object) produces the bulk of the \gray emission above 100 MeV.

We found that the \gray spectrum of \mbox{Cygnus X-3} detected by the \grid is significantly harder than the time-averaged spectrum obtained by \lat for the ``\gray active periods'' of the microquasar, lasting $\sim$4 months (see Figure \ref{cyg_x3_agile_fermi_only}). Although both the AGILE main \gray events and the Fermi \gray active periods are both likely related to the presence of an active jet, the spectral difference may imply that there was a fast hardening of the spectrum during the peak \gray events, lasting $\sim$1-2 days.

We have demonstrated that both a leptonic model based on inverse Compton emission from a relativistic plasmoid injected into the jet and a hadronic model based on $\pi^0$-decays, might account for the \gray emission observed by the \grid. Both of these models require the introduction of a new component (``IC bump'' or ``$\pi^0$-bump'') into the SED of the system. In both the leptonic and hadronic pictures, the inclination of the jet to the line of sight is assumed to be $i = 14^{\circ}$.\\
A leptonic scenario seems to be more likely than a hadronic one: the \gray modulation, the spectral link between hard X-ray and \gray spectra, and the temporal link between \gray events and radio flares could be interpreted in a natural way by assuming that the electrons are the main emitters. According to our results, the HE \gray emission occurs at distances up to $\sim$$ 10^{12}$ cm from the compact object. If we were to interpret the $\sim$4-day delay between the onset of \gray and radio flaring emission as the propagation time of the relativistic jet ($v = \sqrt{5}c/3$), the radio burst would occur at a distance of $\sim$$8\times10^{15}$ cm.\\
Our hadronic model, with the assumption of a standard WR wind, would require a jet kinetic power of $L_{kin,~p} \approx 1.5 \times 10^{38}$ $\mathrm{erg~s^{-1}}$ to explain the \gray emission detected by AGILE. This value is of the same order of magnitude as the bolometric luminosity of the disk/corona during the hypersoft spectral state, and lower than the Eddington accretion limit for a black hole with a mass of $M_x \approx 10 M_{\odot}$ ($L_{Edd} \approx 10^{39}$ $\mathrm{erg~s^{-1}}$). Thus, a hadronic picture is physically reasonable and not energetically less likely than a leptonic one. At present, there is no strong evidence that one of these hypotheses can be excluded, and it remains an open question whether the dominant process for \gray emission in microquasars is either hadronic or leptonic \citep{mirabel_12}. 

The firm discovery of \gray emission from this microquasar represents the experimental proof that these astrophysical objects are capable of accelerating particles up to relativistic energies, through a mechanism -- related to the disk-corona dynamics -- that leads to jet formation.

\begin{acknowledgements}
The authors are grateful to the anonymous referee for her/his stimulating comments on the manuscript.
We also thank A. Zdziarski for discussions about this work.\\
This investigation was carried out with partial support under ASI contracts nos. I/089/06/2, and I/042/10/0.
\end{acknowledgements}

\begin{appendix}
\section{The \grid dataset} \label{agile_data}

We performed an analysis of the whole \grid data in the period November 2007 - July 2009 using a detection algorithm developed by the AGILE team to automatically search for transient \gray emission. The algorithm initially analyzed 140 maps, each related to a 2-day integration (non-overlapping consecutive time intervals). The time bins containing the peak \gray emission, with detection significances greater than 3$\sigma$, were identified. The analysis was subsequently manually refined to optimize the determination of the time interval of the \gray emission. The whole analysis was carried out with the \verb+Build 19+ version of the AGILE team software, using the \verb+FM3.119_2+ calibrated filter applied to the consolidated dataset with off-axis angles smaller than $40^{\circ}$.
We used a multi-source maximum-likelihood analysis (MSLA) to take into account the emission of the nearby \gray pulsars\footnote{The main characteristics of the persistent \gray sources that we used in the MSLA are reported in Table 1 of \citet{chen_piano_11}.} \hp (\hpsr), \gcyg (\gcygpsr), and \srcyg (\srcygpsr). In particular, the MSLA is fundamental to avoid contamination by the pulsar \srcygpsr, located at a distance of $\sim 0.5^{\circ}$. In this paper, we did not consider the off-pulse data for the nearby pulsar. Nevertheless, the MSLA accounted for the steady \gray emission from this source when calculating the significance and the flux of each \gray detection of \mbox{Cygnus X-3}. Moreover, we can exclude any substantial spectral contamination from the pulsar because the steady \gray emission from the pulsar\footnote{The steady \gray emission from the pulsar \srcygpsr, as detected by the \grid, is $F_{\gamma}^{PSR} = [37 \pm 4(stat) \pm 10\% (syst)] \times 10^{-8}$ \flx for photon energies above 100 MeV, see \citet{chen_piano_11} for details.} is much fainter than the mean flux of the active \gray emission from \mbox{Cygnus X-3}.\\
The main events of \gray activity, detected with a significance above $3\sigma$ ($\sqrt{TS} \geqslant 3$), are shown in Table \ref{cyg_x3_all_flares}.
We found seven events\footnote{All the \gray events were detected by using the same filter, FM3.119\_2. As discussed in the supplementary information of \citet{tavani_09a}, the event of 2-3 November 2008 appears to be relatively ``soft'' in \grays compared to the other episodes. By analyzing the event with the FT3ab\_2 filter (which is more efficient in detecting this kind of emission), we found a more significant detection of $\sqrt{TS}=3.9$, at photon fluxes above 100 MeV equal to $(214 \pm 73) \times 10^{-8}$ \flx.}, including those presented in \citet{tavani_09a} and \citet{bulgarelli_12a}.

\begin{table*}
\begin{center}
\caption{Main events of \gray emission detected by the \grid in the period November 2007 - July 2009. All detections have a significance above $3\sigma$ ($\sqrt{TS} \geqslant 3$). \textit{Column 1}: period of detection in UTC; \textit{Column 2}: period of detection in MJD; \textit{Column 3}: significance of detection; \textit{Column 4}: photon flux (above 100 MeV).} \label{cyg_x3_all_flares}

\renewcommand{\arraystretch}{1.5}
\begin{tabular}{|c|c|r|c|}
\hline\hline
Period                   &          MJD        &  $\sqrt{\mathrm{TS}}$  &  Flux [$10^{-8}$ photons $\mathrm{cm^{-2}}$ $\mathrm{s^{-1}}$] \\
\hline
2008 Feb 11 (18:07:28) - 2008 Feb 12 (11:07:44)  & 54507.76 - 54508.46 &        3.7         &           264  $\pm$  104   \\
\hline                                                                                 
2008 Apr 16 (13:59:12) - 2008 Apr 17 (13:48:00)  & 54572.58 - 54573.58 &        4.5         &           265  $\pm$   80   \\
\hline
2008 Nov 2 (13:01:05)  - 2008 Nov 3 (19:01:05)   & 54772.54 - 54773.79 &        3.1         &           135  $\pm$   56    \\
\hline
2008 Dec 11 (19:50:40) - 2008 Dec 12 (23:02:40)  & 54811.83 - 54812.96 &        4.0         &           190  $\pm$   65   \\
\hline
2009 Jun 20 (21:04:48) - 2009 Jun 21 (20:53:04)  & 55002.88 - 55003.87 &        3.8         &           193  $\pm$   67   \\
\hline
2009 Jul 13 (01:11:60) - 2009 Jul 14 (00:59:44)  & 55025.05 - 55026.04 &        3.2         &           216  $\pm$   89   \\
\hline
2009 Jul 21 (21:07:12) - 2009 Jul 23 (21:07:12)  & 55033.88 - 55035.88 &        3.6         &           158  $\pm$   59   \\
\hline
\end{tabular}
\end{center}

\end{table*}

\begin{figure}[!h]
 \begin{center}
    \includegraphics[width=7.5cm]{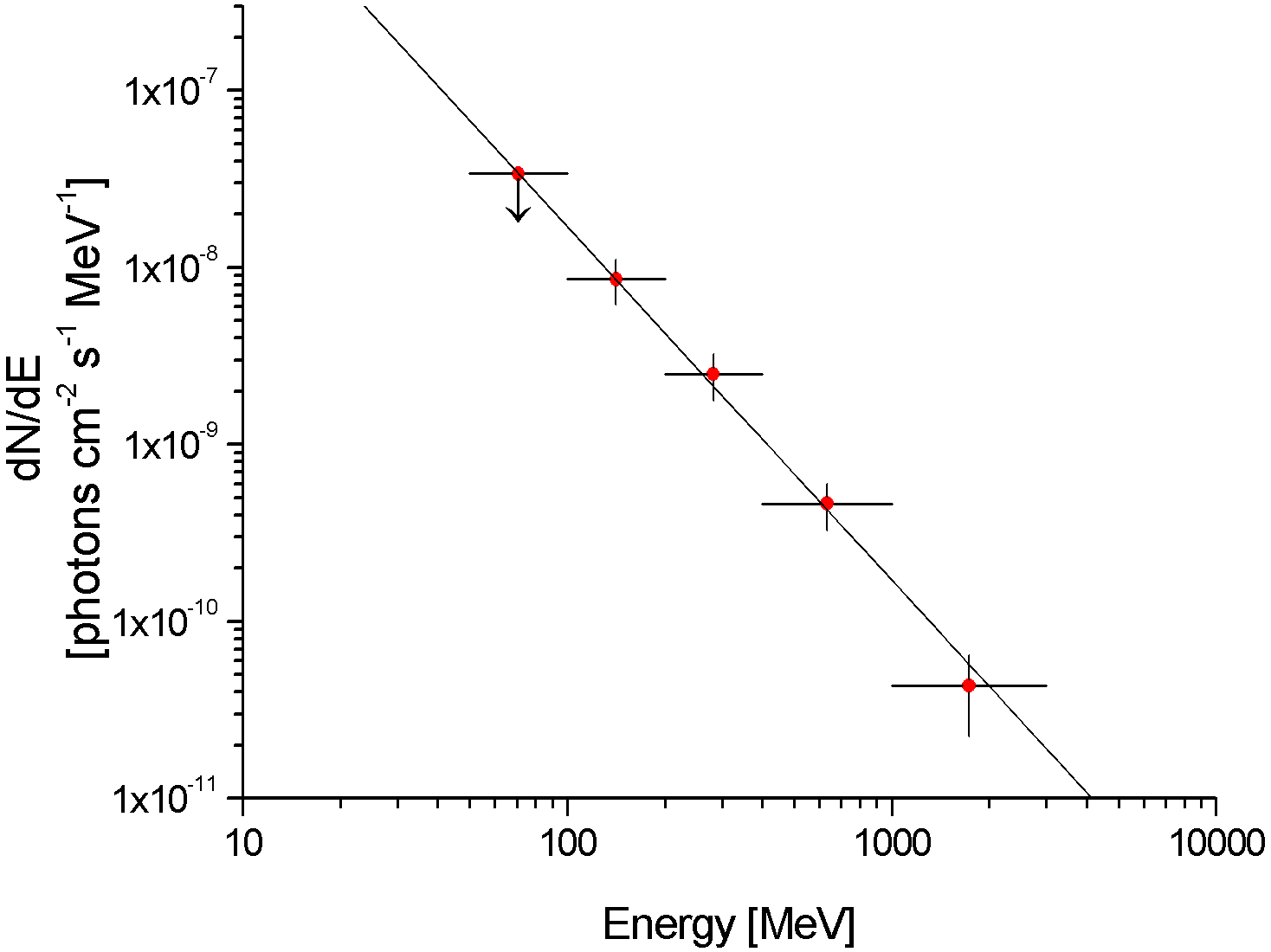}
    \caption{Photon spectrum between 50 MeV and 3 GeV of \mbox{Cygnus X-3} found by the \grid by integrating all the main \gray episodes in Table \ref{cyg_x3_all_flares}. Power-law fit to \gray data between 100 MeV and 3 GeV with photon index $\alpha=2.0 \pm 0.2$.} \label{cyg_x3_spectrum_b20}
\end{center}
\end{figure}

\begin{figure}[!h]
 \begin{center}
    \includegraphics[width=7.5cm]{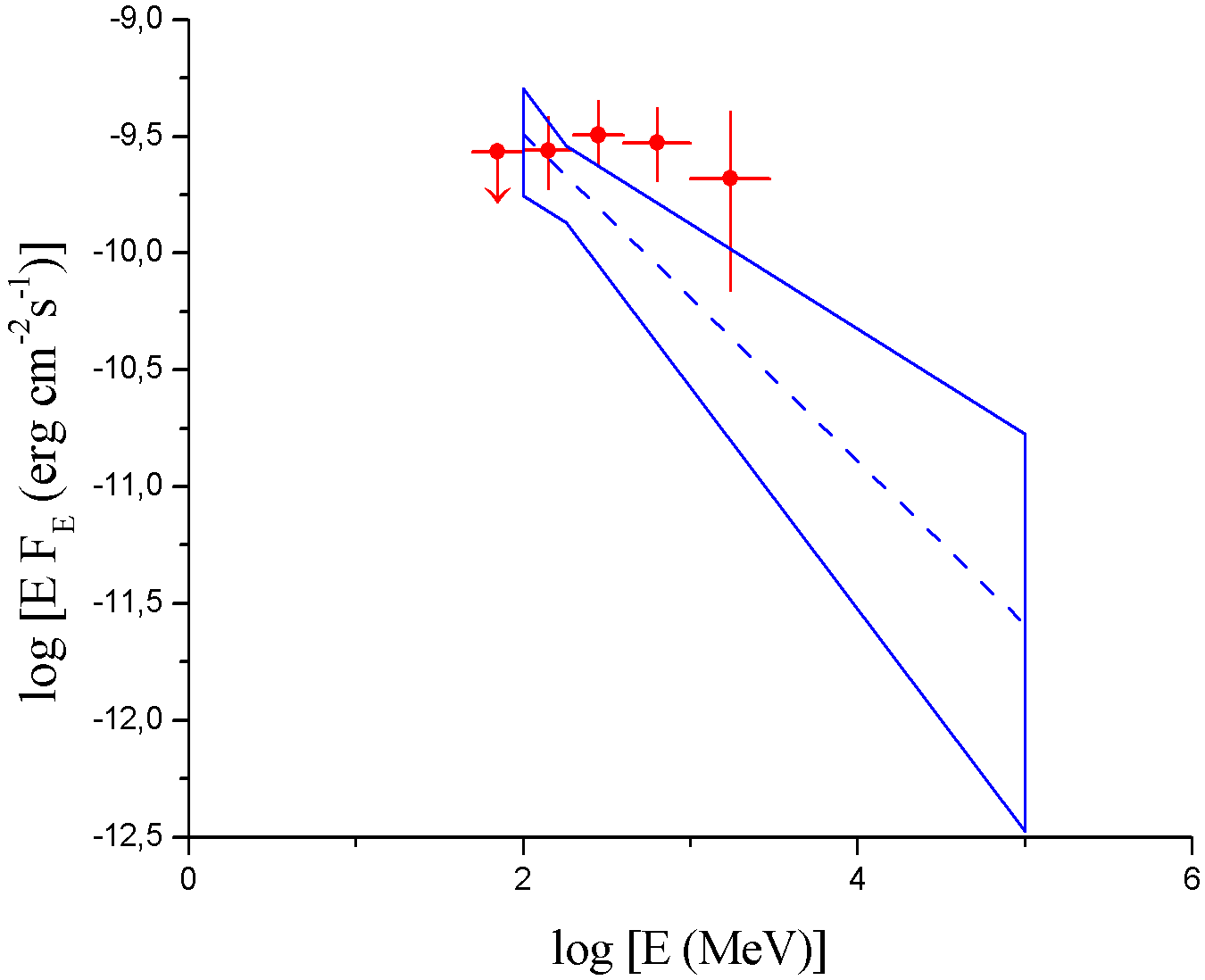}
    \caption{The $\nu F_{\nu}$ spectra of \mbox{Cygnus X-3} during the \gray activity. \textit{Red circles}: \grid energy spectrum (50 MeV to 3 GeV) of the main episodes (Figure~\ref{cyg_x3_spectrum_b20}). \textit{Blue error contours} and \textit{dashed blue line}: average power-law fit with $\alpha=2.70 \pm 0.25$ of the spectrum obtained by \lat integrating the two active windows of about two months each \citep{abdo_09}.} \label{cyg_x3_agile_fermi_only}
\end{center}
\end{figure}

By integrating all the main events with the \verb+FM3.119_2+ filter, we detected a \gray source at $6.7\sigma$ ($\sqrt{TS}=6.7$) at the average Galactic coordinate $(l,~b) = (79.7^{\circ},~0.9^{\circ})$ $\pm~0.4^{\circ}$ (stat) $\pm~0.1^{\circ}$ (syst), with a photon flux of $(158 \pm 29) \times 10^{-8}$ \flx above 100 MeV\footnote{Here we present an updated result for the analysis of the 7-event integration with respect to the one reported in \citet{piano_11}. In this paper, our analysis was carried out with a more recent version of the AGILE software tool (AG\_multi4).}. The average differential spectrum between 100 MeV and 3 GeV is well-fitted by a power law with a photon index $\alpha=2.0~\pm~0.2$ (Figure~\ref{cyg_x3_spectrum_b20}). This value is consistent with the \mbox{Cygnus X-3} photon index found by \citet{tavani_09a} and \citet{bulgarelli_12a}.
In Figure~\ref{cyg_x3_agile_fermi_only}, we compare the $\nu F_{\nu}$ spectra of \mbox{Cygnus X-3} obtained by the \grid and \lat \citep{abdo_09} during the \gray activity. We remark that the \grid spectrum is related only to the peak \gray activity (the seven main events, lasting 1-2 days, in Table \ref{cyg_x3_all_flares}), whereas the \lat spectrum is an average spectrum found during the two active windows (of about two months each) of \gray emission from \mbox{Cygnus X-3} (MJD: 54750--54820 and MJD: 54990--55045).

Finally, we evaluated the post-trial significance for repeated flare occurrences by using the same formalism of \citet{bulgarelli_12b}. The probability of having $k$ or more detections -- consistent with the position of \mbox{Cygnus X-3} -- with $\sqrt{TS} \geqslant \sqrt{h}$ in $N$ trials, is
$$
P(N,k) = 1 - \sum_{j=0}^{k-1} \dbinom{N\,}{j\,} \; p^{j} (1-p)^{N-j} ~,
$$
where $p$ is the $p$-value corresponding to $h$. For $\sqrt{TS} \geqslant 3$, we have a $p$-value of $p = 2.0 \times 10^{-3}$. Thus, for $N=140$ (our trials, the number of 2-day integration maps analyzed by the initial algorithm) and $k=7$ (our detections of \mbox{Cygnus X-3}), we found $P(140,7)=1.8\times10^{-8}$, which corresponds to 5.5 Gaussian standard deviations.

An MSLA applied to the deep integration of the \grid data (between November 2007 and July 2009) found weak persistent emission from a position consistent with \mbox{Cygnus X-3}\footnote{The persistent \grid source associated to \mbox{Cygnus X-3} is J2033+4050 in Table 1 of \citet{chen_piano_11}} (significance $\sqrt{TS}=5.17$ and photon flux $F_{\gamma} = (14 \pm 3) \times 10^{-8}$ \flx).

\end{appendix}

\end{document}